# A New Hybrid JPEG Image Compression Scheme Using Symbol Reduction Technique

Bheshaj Kumar[1],Kavita Thakur[2] and G. R. Sinha[3]

[1,2]School of Studies in Electronics, Pt. R. S. University,Raipur,India
bheshaj.dewangan@rediffmail.com
kavithakur@rediffmail.com
[3]Shri Shankarachrya College of Engineering & Technology, Bhilai, India
ganeshsinha2003@gmail.com

## *Abstract*

*Lossy JPEG compression is a widely used compression technique. Normally the JPEG standard technique uses three process mapping reduces interpixel redundancy, quantization, which is lossy process and entropy encoding, which is considered lossless process. In this paper, a new technique has been proposed by combining the JPEG algorithm and Symbol Reduction Huffman technique for achieving more compression ratio. The symbols reduction technique reduces the number of symbols by combining together to form a new symbol. As a result of this technique the number of Huffman code to be generated also reduced. It is simple fast and easy to implement. The result shows that the performance of standard JPEG method can be improved by proposed method. This hybrid approach achieves about 20% more compression ratio than the Standard JPEG.*

## 1. Introduction

Image compression has a fundamental importance in image communication and storage. A typical image compression scheme first manipulates the input image data in a way to obtain more compact and/or uncorrelated representation. This manipulation may be in the form of a mapping, transformation, quantization or a combination of these. The manipulated data is then fed to an entropy encoder [1-2]. Recent research in the field of image compression focuses more on data manipulation stage, and a growing number of new techniques concerning this stage are introduced in the recent year. Conventional technique used in the entropy encoding stage and there are relatively less innovation in this stage. Thus the performance of most image compression scheme can be improved by utilizing an entropy encoder.

In most general terms, an entropy coder is an invertible mapping from a sequence of events to a sequence of bits. The objective of entropy coding is the minimization of the number of bits in the bit sequence, while maintaining the invertibility of the mapping. The way of defining the particular events can also be considered a part of entropy coding. The entropy coding can be considered to define its own set of intermediate events, based on the input sequence of events as define externally. Entropy coding has been extensively studies in the literature, particularly under the discipline of information theory.

Let $X = x_1x_2x_3\ldots\ldots\ldots\ldots x_n$ be a sequence of independent input symbols, each symbol taking on values from a finite alphabet $A = \{ a_1,a_2,\ldots\ldots\ldots a_m)$. Let $p_{ij}$ denote the probability that $x_i$, the $i^{th}$ symbol in the sequence, takes the value $a_j$. Thus the entropy of the sequence is defined as

$$H(X) = -\sum_{i=1}^{n}\sum_{j=1}^{m} p_{ij} \log_2 p_{ij} \qquad (1)$$

Shannon's source coding theorem [1] states that the entropy of the sequence is the lower bound on the expected value of the number of bits that is required to encode the sequence [2].

It's well known that the Huffman's algorithm is generating minimum redundancy codes compared to other algorithms [3-7]. The Huffman coding has effectively used in text, image, video compression, and conferencing system such as, JPEG, MPEG-2, MPEG-4, and H.263 etc. .The Huffman coding technique collects unique symbols from the source image and calculates its probability value for each symbol and sorts the symbols based on its probability value. Further, from the lowest probability value symbol to the highest probability value symbol, two symbols combined at a time to form a binary tree. Moreover, allocates zero to the left node and one to the right node starting from the root of the tree. To obtain Huffman code for a particular symbol, all zero and one collected from the root to that particular node in the same order [8-11].

There are numerous amount of work on image compression is carried out both in lossless and lossy compression [11-14]. Very limited research works are carried out for Hybrid Image compression[21-27]. Hsien Wen Tseng and Chin-Chen Chang proposed a very low bit rate image compression scheme that combines the good energy-compaction property of DCT with the high compression ratio of VQ-based coding[15]. Lenni Yulianti and Tati R.Mengko proposed a hybrid method to improve the performance of Fractal lmage Compression (FIC) technique by combining FIC (lossy compression) and lossless Huffman coding[16].A novel hybrid image compression technique for efficient storage and delivery of data is proposed by S.Parveen Banu and Dr.Y.Venkataramani based on decomposing the data using daubechies-4 wavelet in combination with the lifting scheme and entropy encoding[17].

In the proposed work, for image compression, symbol reduction technique is applied in Standard JPEG lossy method. The results provide better compression ratio compare to JPEG technique.

# 2. BACKGROUND OF OUR WORK

## 2.1. Joint Photographic Expert Group (JPEG)

A common characteristic of most images is that the nearby pixels are associated and therefore comprise redundant information. The primary task then is to find less associated representation of the image. Two essential attributes of compression are redundancy and irrelevancy reduction. Redundancy reduction aims at removing repetition from the signal source. Irrelevancy reduction omits parts of the signal that will not be discerned by the signal receiver, namely the Human Visual System (HVS). JPEG Compression method is one of the commonly recognized and used procedures for image compression. JPEG stands for Joint Photographic Experts Group. JPEG standard has been established by ISO (International Standards Organization) and IEC (International Electro-Technical Commission) [18]. The JPEG Image compression system consists of three closely connected constituents specifically

- Source encoder (DCT based)
- Quantizer
- Entropy encoder [18]

Figure 1shows the block diagram of a JPEG encoder, which has the following components [18]

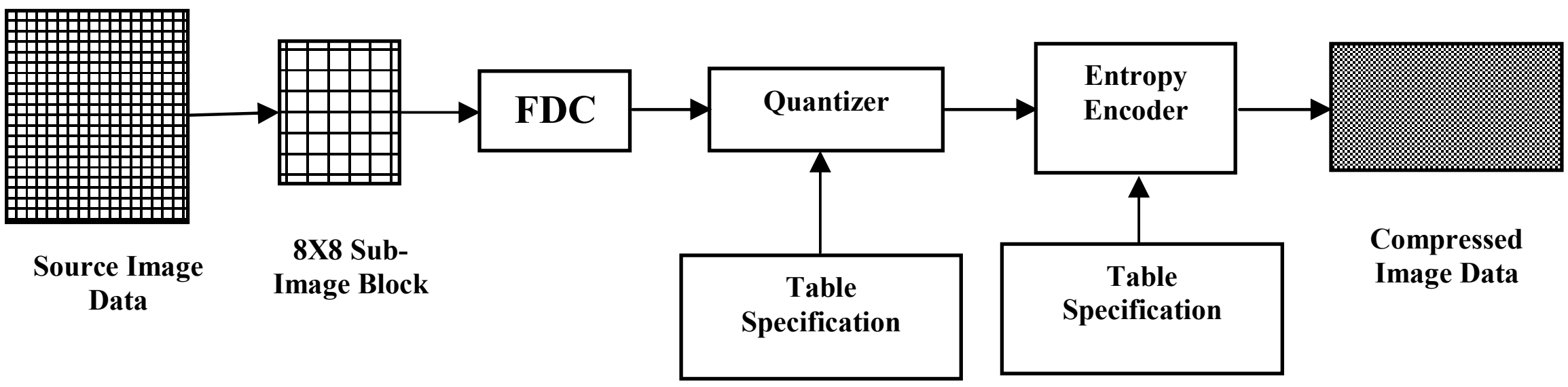

Figure 1. Block Diagram of JPEG Encoder [18]

### 2.1.1 Forward Discrete Cosine Transform (FDCT)

The still images are first partitioned into non-overlapping blocks of size 8x8 and the image samples are shifted from unsigned integers with range [0 to $2^p$-1] to signed integers with range [-$2^{p-1}$ to $2^{p-1}$ ], where $p$ is the number of bits. It should however be mentioned that to preserve freedom for innovation and customization within implementations, JPEG neither specifies any unique FDCT algorithm, nor any unique IDCT algorithms[18]. Because adjacent image pixels are highly correlated, the `forward' DCT (FDCT) processing step lays the foundation for achieving data compression by concentrating most of the signal in the lower spatial frequencies. In principle, the DCT introduces no loss to the source image samples; it merely transforms them to a domain in which they can be more efficiently encoded [20]. The 2D Discrete Cosine Transform Pair in two dimensions, for a square matrix, is given by

$$F(i,j) = \propto(i) \propto(j) \sum_{M=0}^{N-1} \sum_{N=0}^{N-1} f(m,n) \cos\frac{\pi(2m+1)i}{2N} \cos\frac{\pi(2n+1)j}{2N} \quad (2)$$

For i= 0,1,2,…….,N-1 and j=0,1,2,…………,N-1

$$f(m,n) = \sum_{i=0}^{N-1} \sum_{j=0}^{N-1} \propto(i) \propto(j) F(i,j) \cos\frac{\pi(2m+1)i}{2N} \cos\frac{\pi(2n+1)j}{2N} \quad (3)$$

For m= 0,1,2,…….,N-1 and n=0,1,2,…………,N-1

Where,

$$\propto(k) = \begin{cases} \sqrt{1/N} \\ \sqrt{2/N} \end{cases}$$

$$\propto(k) = \begin{cases} \sqrt{1/N}, & for\ k = 0 \\ \sqrt{2/N}, & for\ k = 1,2,\ldots\ldots,N-1 \end{cases}$$

### 2.1.2 Quantization

After output from the FDCT, each of the 64 DCT coefficients block is uniformly quantized according to a quantization table. Since the aim is to compress the images without visible artifacts, each step-size should be chosen as the perceptual threshold or for "just noticeable distortion". Psycho-visual experiments have led to a set of quantization tables and these appear in ISO-JPEG standard as a matter of information, but not a requirement [18]. At the decoder, the quantized values are multiplied by the corresponding QT elements to recover the original unquantized values. After quantization, all of the quantized coefficients are ordered into the *zigzag* sequence. This ordering helps to facilitate entropy encoding by placing low-frequency non-zero coefficients before high-frequency coefficients. The DC coefficient, which contains a significant fraction of the total image energy, is differentially encoded [20].

### 2.1.3 Entropy Coder

This is the final processing step of the JPEG encoder. Entropy Coding (EC) achieves additional compression lossless by encoding the quantized DCT coefficients more compactly based on their statistical characteristics. The JPEG standard specifies two entropy coding methods – Huffman and arithmetic coding. The baseline sequential JPEG uses Huffman only, but codecs with both methods are specified for the other modes of operation. Huffman coding requires that one or more sets of coding tables are specified by the application. The same table used for compression is used needed to decompress it. The baseline JPEG uses only two sets of Huffman tables – one for DC and the other for AC. Figure 2 shows the block diagram of the JPEG decoder. It performs the inverse operation of the JPEG encoder [18].

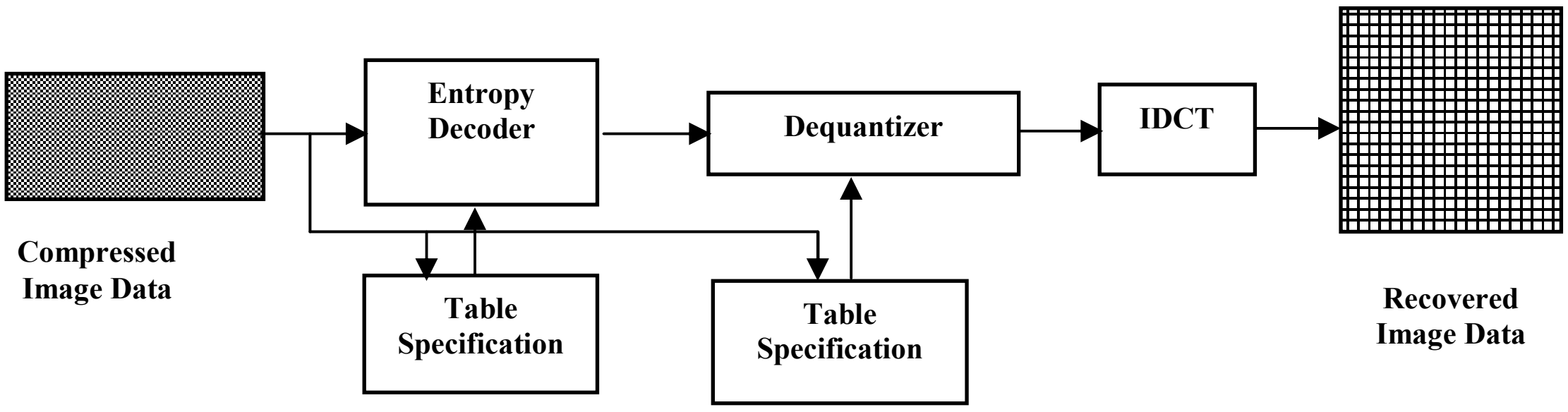

Figure 2. Block Diagram of JPEG Decoder

The use of uniformly sized blocks simplified the compression system, but it does not take into account the irregular shapes within the real images [20]. Degradation occurs which is known as blocking effect and it depends on the block size. A larger block leads to more efficient coding but requires more computational power. Image distortion is less annoying for small than for large DCT blocks. Therefore more existing systems use blocks of 8x8 or 16x16 pixels as a compromise between coding efficiency and image quality [28-29].

## 2.2 Symbol Reduction Method

The number of source symbols is a key factor in achieving compression ratio. A new compression technique proposed to reduce the number of source symbols. The source symbols combined together in the same order from left to right to form a less number of new source symbols. The source symbols reduction explained with an example as shown below. The following eight symbols are assumed as part of an image, A, B, C, D, E, F, G, H. By applying source symbols reduction from left to right in the same sequence, four symbols are combined together to form a new element, thus two symbols ABCD and EFGH are obtained[8]. This technique helps to reduce 8 numbers of source symbols to 2 numbers i.e. $2^n$ symbols are reduced to $2^{(n-2)}$ symbols. For the first case, there are eight symbols and the respective Symbols and Huffman Codes are A-0, B-10, C-110, D-1110, E-11110, F-111110, G-1111110, H-1111111. The proposed technique reduced the eight symbols to two and the reduced Symbols and Huffman codes are ABCD-0, EFGH-1.

The average number $L_{avg}$ of bits required to represent a symbol is defined as,

$$.L_{avg} = \sum_{k=1}^{l} l(r_k)p(r_k) \qquad (4)$$

where, $r_k$ is the discrete random variable for k=1,2,…L with associated probabilities $p_r(r_k)$. The number of bits used to represent each value of $r_k$ is $l(r_k)$ [9]. The number of bits required to represent an image is calculated by number of symbols multiplied by $L_{avg}$ [1-2].In the Huffman coding, probability of each symbols is 0.125 and $L_{avg}$ = 4.375.In the proposed technique, probability of each symbol is 0.5 and $L_{avg}$ =1.0.The $L_{avg}$ confirms that the proposed technique achieves better compression than the Huffman Coding.

8 rows and 8 columns of eight bits grey-scale image having 64 symbols considered to calculate required storage size. In the coding stage these two techniques make difference [9]. In the first case, 64 symbols generate 64 Huffman codes, whereas the proposed technique generates 16 Huffman codes and reduces $L_{avg}$. Therefore, the experiment confirms that the source symbols reduction technique helps to achieve more compression [9].

Since the source images firstly divided in 8x8 sub-block, will have totally 64 symbols. The images are 8 bit grey-scale and the symbol values range from 0 to 255. To represent each symbol eight bit is required [9]. Therefore, size of an image becomes 64 x 8 = 512 bit.

# 3. PROPOSED JPEG IMAGE COMPRESSION METHOD

The image is subdivided into non-overlapping 8x8 sub-image blocks and DCT coefficients are computed for each block. The quantization is performed conferring to quantization table. The quantized values are then rearranged according to zigzag arrangement. After getting zigzag coefficients the remaining coefficients are compressed by the proposed entropy encoder. The block diagram of our proposed method is shown in figure 3.

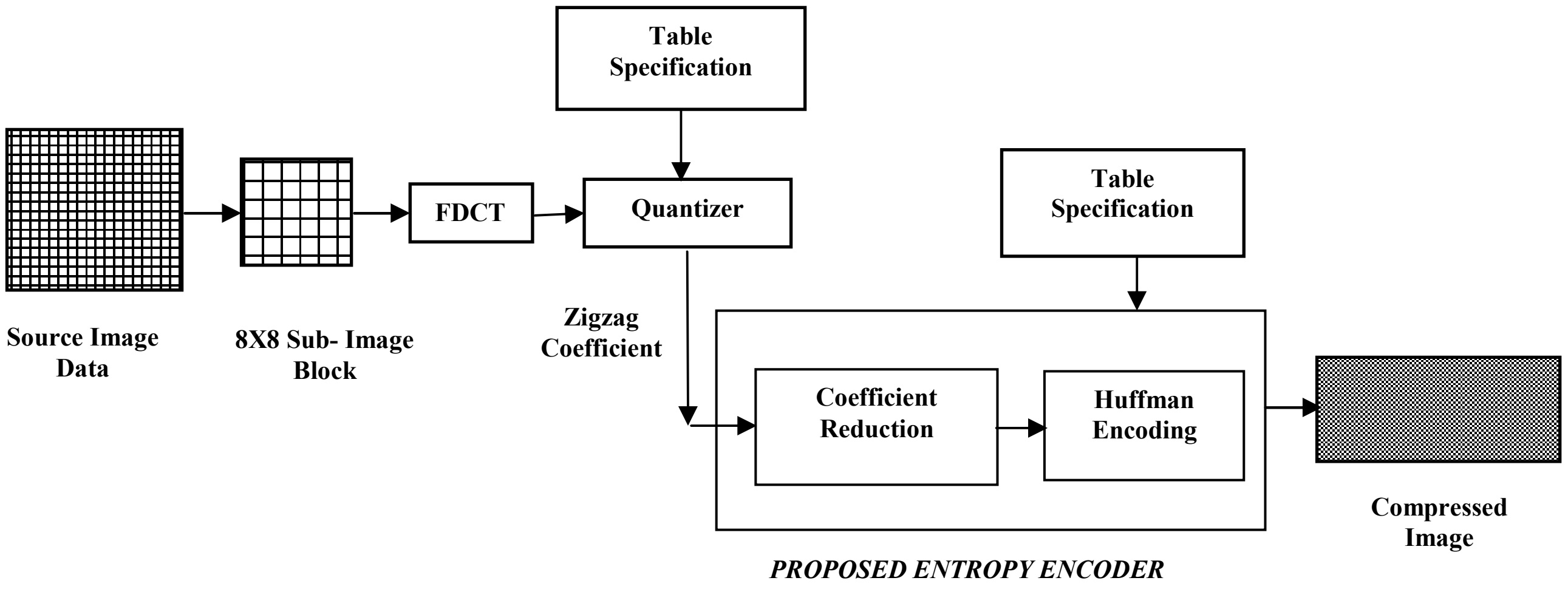

Figure 3. Block Diagram of Proposed JPEG Encoder

## 3.1 Algorithm

1. Input the image data to be compressed.
1. Divide the image into non-overlapping 8x8 sub images blocks.

2. Shift the gray-levels in the range between [-128, 127].
3. Apply DCT on the each sub-image.
4. Quantize the coefficients and the less significant coefficients are set to zero.
5. Order the coefficients using zigzag ordering and the coefficients obtained are in order of increasing frequency.
6. Compress remaining quantized values by applying proposed entropy encoder.

To reconstruct the image, reverse process of our proposed algorithm are carried out.

As an example, let us consider the 8 X 8 block of pixel value component samples shown in Table 1. This block is extracted from one of the images used in the experiments. Table 2 shows the corresponding coefficients obtained by applying the DCT transform to the block in Table 1.

Table 1. An example of pixel values of 8X8 image block

| | | | | | | | |
|---|---|---|---|---|---|---|---|
| 58 | 45 | 29 | 27 | 24 | 19 | 17 | 20 |
| 62 | 52 | 42 | 41 | 38 | 30 | 22 | 18 |
| 48 | 47 | 49 | 44 | 40 | 36 | 31 | 25 |
| 59 | 78 | 49 | 32 | 28 | 31 | 31 | 31 |
| 98 | 138 | 116 | 78 | 39 | 24 | 25 | 27 |
| 115 | 160 | 143 | 97 | 48 | 27 | 24 | 21 |
| 99 | 137 | 127 | 84 | 42 | 25 | 24 | 20 |
| 74 | 95 | 82 | 67 | 40 | 25 | 25 | 19 |

Table 2. DCT coefficients corresponding to the 8X8 block in Table 1

| | | | | | | | |
|---|---|---|---|---|---|---|---|
| 421.00 | 203.33 | 10.65 | -45.19 | -30.25 | -13.83 | -14.15 | -7.33 |
| -107.82 | -93.43 | 10.09 | 49.21 | 27.72 | 5.88 | 8.33 | 3.28 |
| -41.83 | -20.47 | -6.16 | 15.53 | 16.65 | 9.09 | 3.28 | 2.52 |
| 55.94 | 68.58 | 7.01 | -25.38 | -9.81 | -4.75 | -2.36 | -2.12 |
| -33.50 | -21.10 | 16.70 | 8.12 | 3.25 | -4.25 | -4.75 | -3.39 |
| -15.74 | -13.60 | 8.12 | 2.42 | -3.98 | -2.12 | 1.22 | 0.73 |
| 0.28 | -5.37 | -6.47 | -0.58 | 2.30 | 3.07 | 0.91 | 0.63 |
| 7.78 | 4.95 | -6.39 | -9.03 | -0.34 | 3.44 | 2.57 | 1.93 |

Now we apply the quantization processes to the DCT coefficients in Table 2. To compute the quantized values, we adopt the JPEG standard default quantization table as shown in Table 3. Table 4 reports the quantized DCT coefficients.

Table 3. Default Quantization Table

| | | | | | | | |
|---|---|---|---|---|---|---|---|
| 16 | 11 | 10 | 16 | 24 | 40 | 51 | 61 |
| 12 | 12 | 14 | 19 | 26 | 58 | 60 | 55 |
| 14 | 13 | 16 | 24 | 40 | 57 | 69 | 56 |
| 14 | 17 | 22 | 29 | 51 | 87 | 80 | 62 |
| 18 | 22 | 37 | 56 | 68 | 109 | 103 | 77 |
| 24 | 35 | 55 | 64 | 81 | 104 | 113 | 92 |
| 49 | 64 | 78 | 87 | 103 | 121 | 120 | 101 |
| 72 | 92 | 95 | 98 | 112 | 100 | 103 | 99 |

Table 4. The quantized DCT coefficients corresponding to the block in Table 1

| | | | | | | | |
|---|---|---|---|---|---|---|---|
| 26 | 18 | 1 | -3 | -1 | 0 | 0 | 0 |
| -9 | -8 | 1 | 3 | 1 | 0 | 0 | 0 |
| -3 | -2 | 0 | 1 | 0 | 0 | 0 | 0 |
| 4 | 4 | 0 | -1 | 0 | 0 | 0 | 0 |
| -2 | -1 | 0 | 0 | 0 | 0 | 0 | 0 |
| -1 | 0 | 0 | 0 | 0 | 0 | 0 | 0 |
| 0 | 0 | 0 | 0 | 0 | 0 | 0 | 0 |
| 0 | 0 | 0 | 0 | 0 | 0 | 0 | 0 |

After quantization, we can observe that there usually are few nonzero and several zero-valued DCT coefficients. Thus, the objective of entropy coding is to losslessly compact the quantized DCT coefficients exploiting their statistical characteristics. The quantized 64 coefficients are formatted by using the zigzag scan preparation for entropy coding.

[ 26 18 -9 -3 -8 1 -3 1 -2 4 -2 4 0 3 -1 0 1 1 0 -1 -1 0 0 0 -1 0 0 0 0 0 0 0 0 0 0 0 0 0 0 0 0 0 0 0 0 0 0 0 0 0 0 0 0 0 0 0 0 0 0 0 0 0 0 0 0]

Since many of the coefficients are negative and zero after quantization, each coefficient is checked for either non-zero or zero coefficient, if coefficient is nonzero, add 128 to the coefficient to make a positive integer.

[154 146 119 125 120 129 125 128 126 132 126 132 0 131 127 0 129 129 0 127 127 0 0 0 127 0 0 0 0 0 0 0 0 0 0 0 0 0 0 0 0 0 0 0 0 0 0 0 0 0 0 0 0 0 0 0 0 0 0 0 0 0 0 0 0]

Now according to proposed reduction method, from left to right in the same sequence, four symbols are combined together to form a new string, thus these 64 symbols are reduced in 16 symbols.

[(154146119125) (120129125128) (126132126132) (01311270) (1291290127) (127000) (127000) (0000) (0000) (0000) (0000) (0000) (0000) (0000) (0000) (0000)]

This string is encoded using a predetermined Huffman coding process and finally, it generates compressed JPEG image data.

## 4. EVALUATION OF PROPOSED METHOD

In carrying out the evaluation as part of this research paper, numerical evaluation is adopted. In the case of lossy compression, however, the reproduction is only an approximation to the original image. Measurement of quality and Compression Ratio are a central issue with lossy compression. Since our main target were to reduce the source symbol after quantization, which is totally reversible process , therefore the compression error not considered. So the picture quality (PSNR) values are common to both Standard JPEG method and Proposed JPEG compression method.So in this paper, we only evaluated the compression ratio ($C_R$ ) using equation 5.

$$C_R = \frac{Original\ Image\ Size}{Compressed\ Image\ Size} \quad (5)$$

# 5. RESULT AND DISCUSSION

For evaluating the performance of the proposed algorithms, we used 256x256 grayscale versions of the well-known Cameraman, Rice, Coin and Tree images, shown in Figure 4. A software algorithm has been developed and implemented to compress the given image using JPEG standard and Huffman coding techniques in a MATLAB platform.

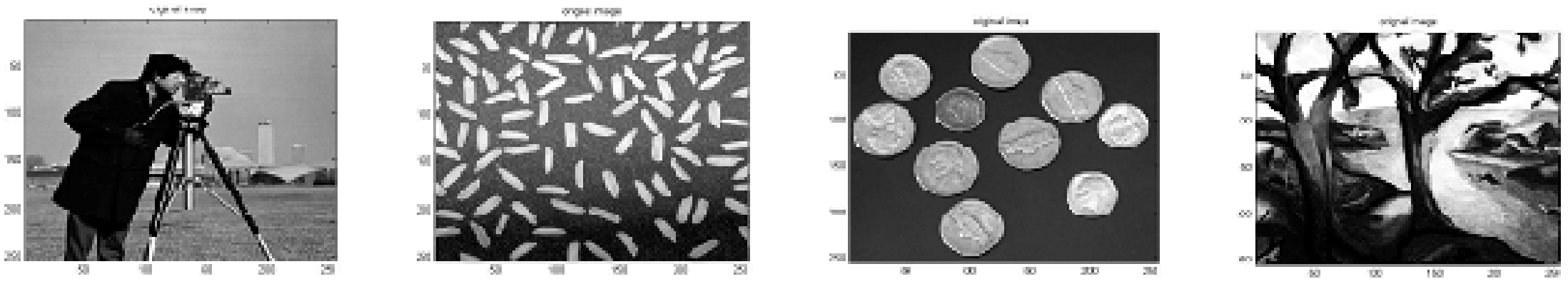

Figure 4. Test images for evaluating the performance of proposed method

Intermediate Images of Cameraman during processing are shown below

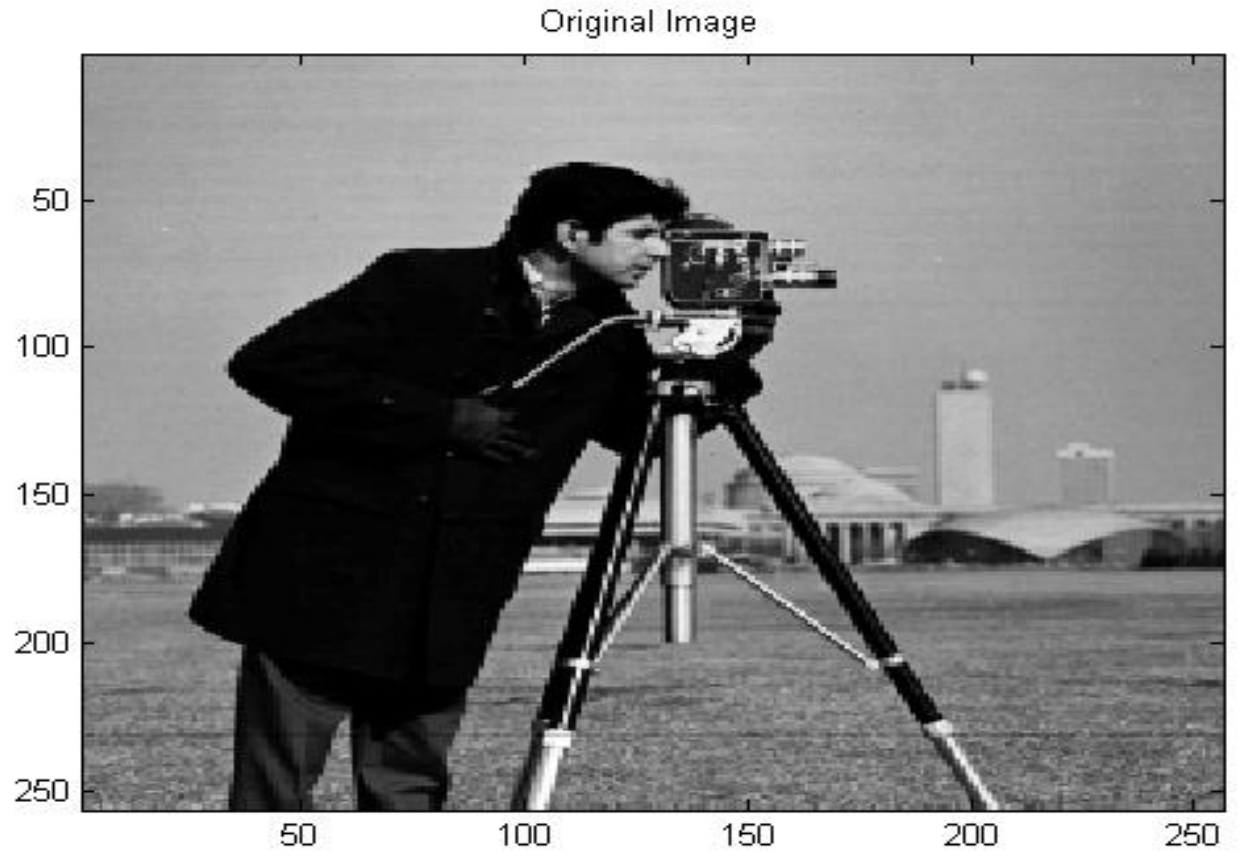

Figure 5. Original Test images

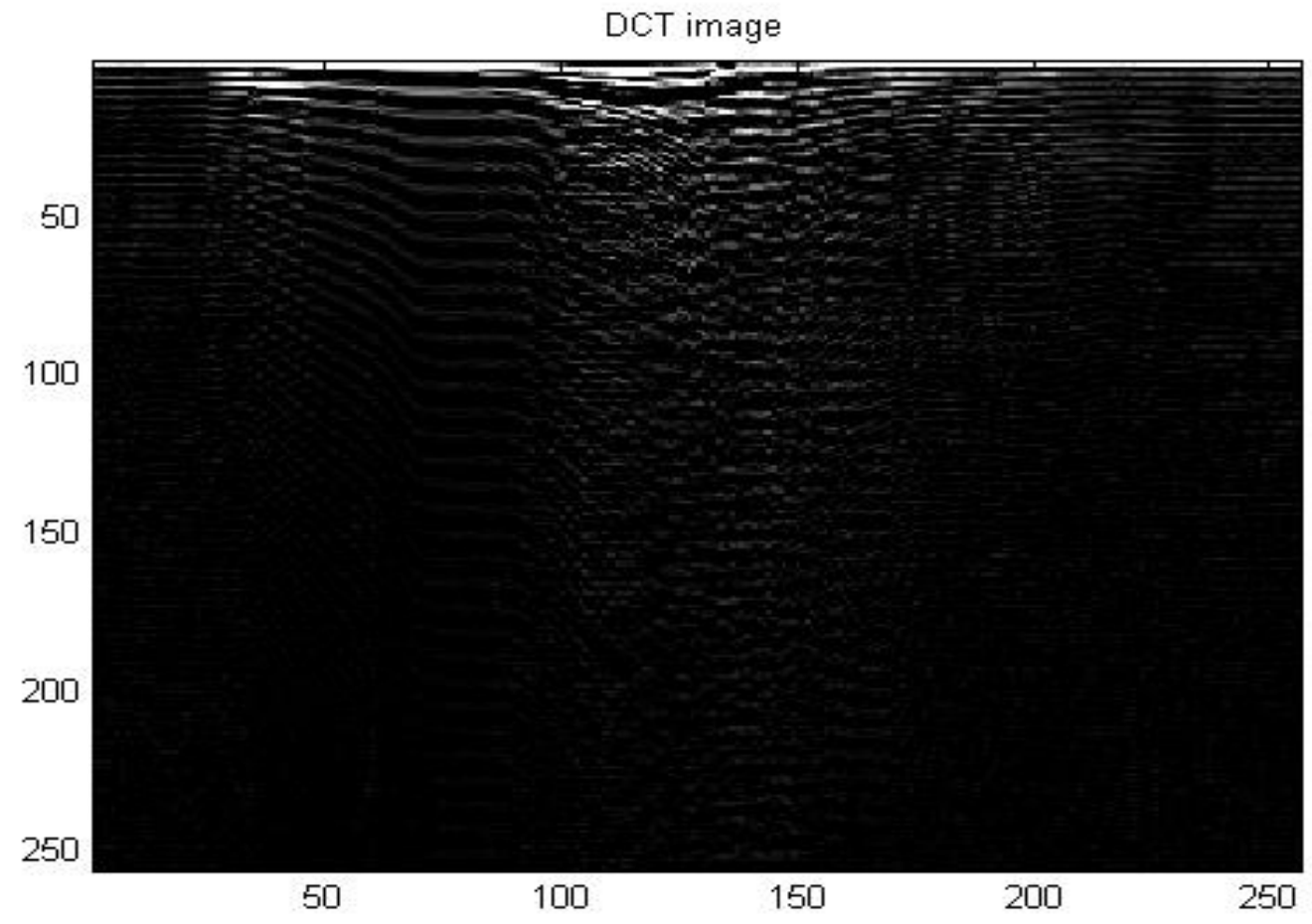

Figure 6. After applying DCT

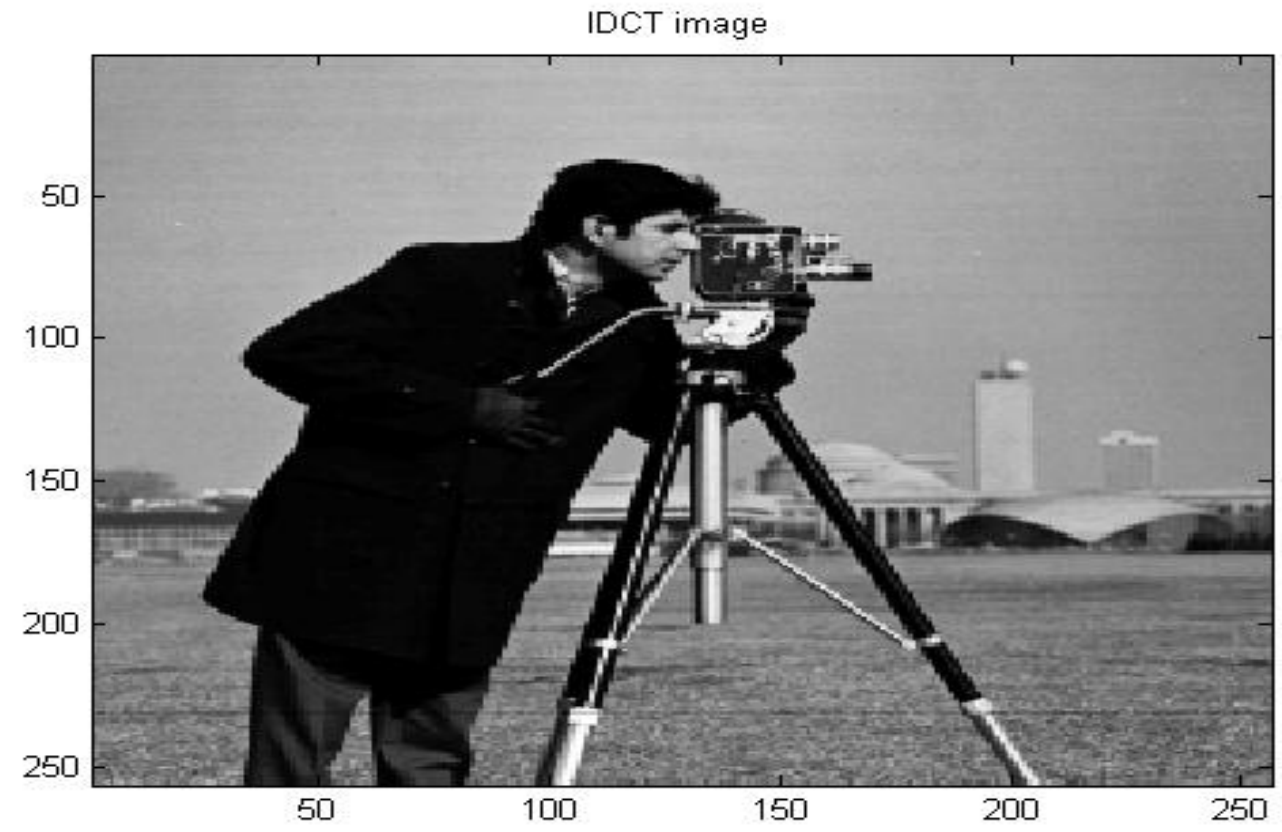

Figure 7. After applying IDCT

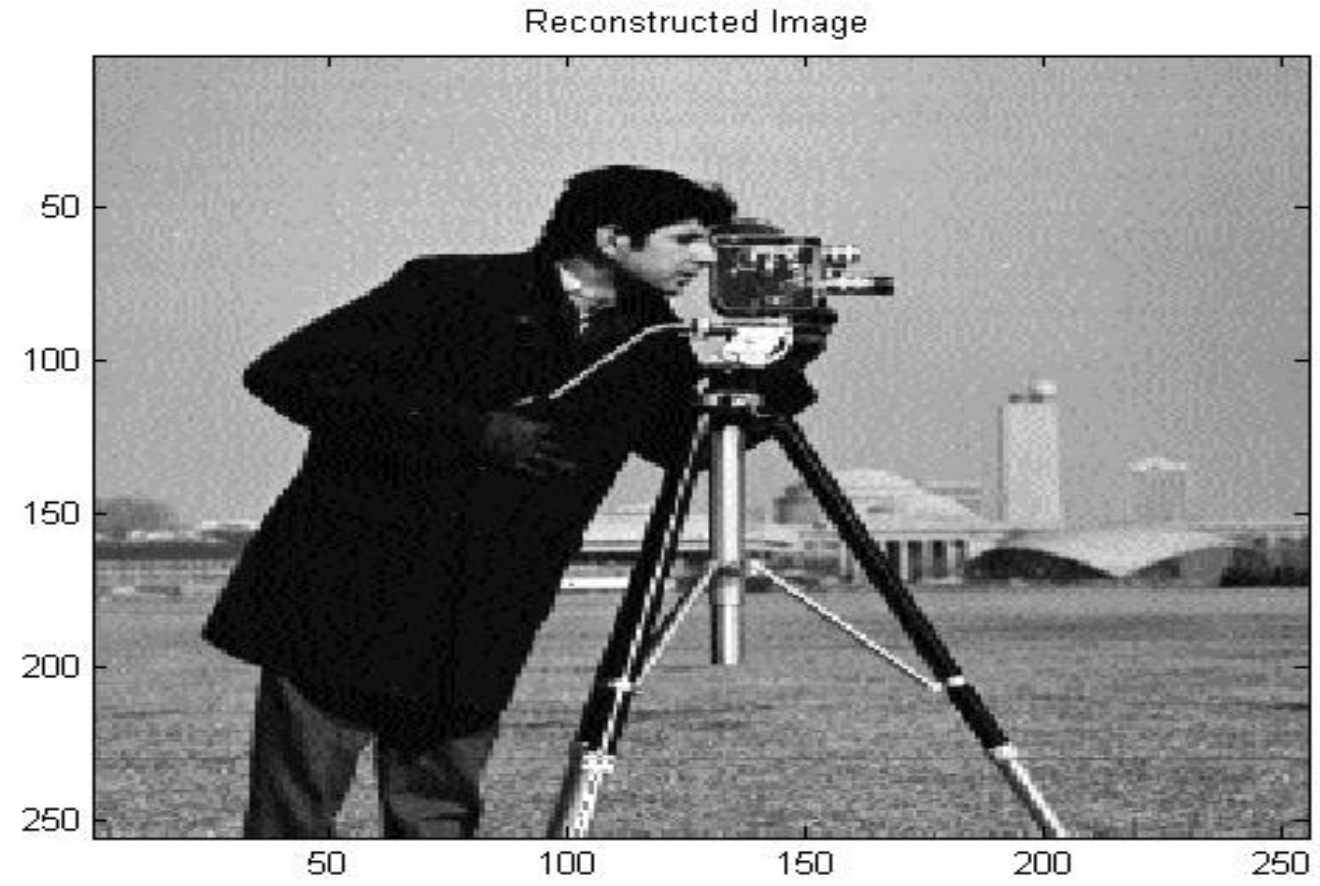

Figure 8. Reconstructed image

The results obtained from the implementation of the proposed algorithm are shown in the table 5. The compressed ratio of the four test images are calculated and presented in table 5.

Table 5. Compression Ratio of Standard JPEG and Proposed Method

| Test Images | JPEG Method | Proposed Method |
|---|---|---|
| Cameraman | 4.13 | 4.94 |
| Rice | 5.68 | 6.45 |
| Coin | 3.84 | 4.73 |
| Tree | 2.53 | 3.08 |

The result shows that for all test images the compressed size obtained from the proposed technique better than the Standard JPEG method. The proposed compression technique achieves better compression. The results obtained from the present analysis are shown in figure 5.

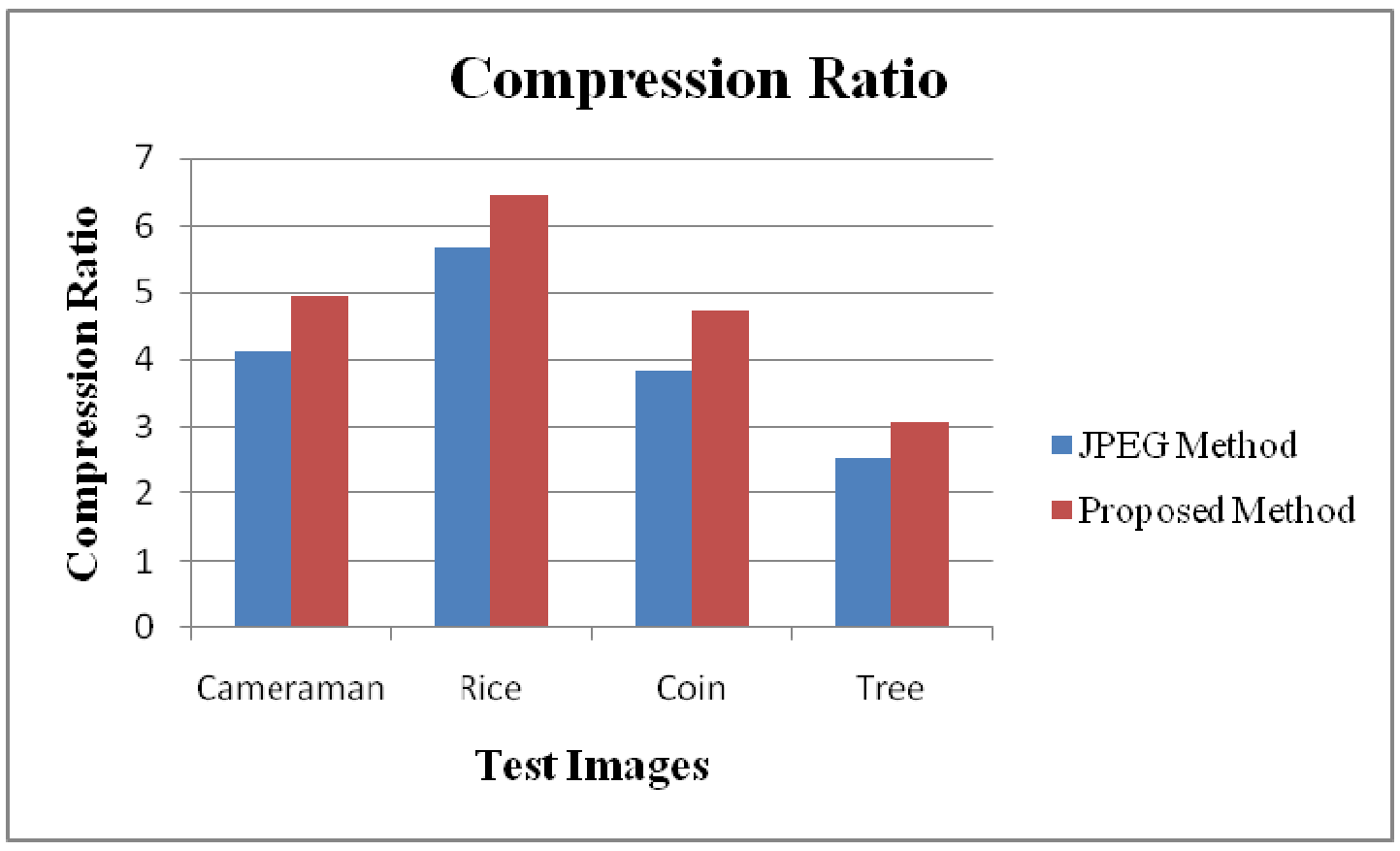

Figure 9. Compression Ratio of Standard JPEG and Proposed Method

Since it is found that the two compression techniques are lossless compression technique, therefore the compression error not considered. So the picture quality (PSNR) values are common to both Standard JPEG method and Proposed JPEG compression method.

## 6. CONCLUSION

The experimentation in present paper reveals that the compression ratio of the proposed technique achieves better than the standard JPEG method. The proposed entropy encoding technique enhances the performance of the JPEG technique. Therefore, the experiment confirms that the proposed technique produces higher lossless compression than the Huffman Coding and this technique will be suitable for compression of all type of image files.

## 7. REFERENCES

[1] C. E. Shannon, “A Mathematical Theory of Communication”, in Bell System Technical Journal, vol. 27, July, October 1948, pp. 379–423,623–656.

[2] Gonzalez, R.C. and Woods, R.E., “Digital Image Processing”, 2nd ed., Pearson Education, India, 2005.

[3] Huffman, D.A., “A method for the construction of minimum-redundancy codes”, Proc. Inst.Radio Eng. 40(9), pp.1098-1101, 1952.

[4] Steven Pigeon, Yoshua Bengio,”A Memory-Efficient Huffman Adaptive Coding Algorithm for Very Large Sets of Symbols”, Université de Montréal, Rapport technique #1081.

[5] Steven Pigeon, Yoshua Bengio,"A Memory-Efficient Huffman Adaptive Coding Algorithm for Very Large Sets of Symbols Revisited" ,Université de Montréal, Rapport technique#1095.
[6] R.G. Gallager,"Variation on a theme by Huffman", IEEE. Trans. on Information Theory, IT-24(6), 1978, pp. 668-674.
[7] D.E. Knuth,"Dynamic Huffman Coding",Journal of Algorithms, 6, 1983 pp. 163-180.
[8] C. Saravanan & R. Ponalagusamy,'Lossless Grey-scale Image Compression using Source Symbols Reduction and Huffman Coding" International Journal of Image Processing , Vol. 3,Issue 5,pp. 246-251.
[9] Jagdish H. Pujar and Lohit M. Kadlaskar, "A New Lossless Method of Image Compression and Decompression Using Huffman Coding Techniques", Journal of Theoretical and Applied Information Technology,pp. 18-23.
[10]Chiu-Yi Chen; Yu-Ting Pai; Shanq-Jang Ruan, "Low Power Huffman Coding for High Performance", Data Transmission, International Conference on Hybrid Information Technology, 2006, 1(9-11), 2006 pp.71 – 77.
[11] Lakhani, G, "Modified JPEG Huffman coding", IEEE Transactions Image Processing, 12(2), 2003 pp. 159 – 169.
[12] K. Rao, J. Hwang, "Techniques and Standards for Image, Video, and Audio Coding", Prentice-Hall, Inc., Upper Saddle River, NJ, USA, 1996.
[13] W.B. Pennebaker, J.L. Mitchell, "JPEG Still Image Data Compression Standard," 2nd ed., Kluwer Academic Publishers, Norwell, MA, USA, 1992.
[14] B. Sherlock, A. Nagpal, D. Monro, "A model for JPEG quantization", Proceedings of the International Symposium on Speech, Image Processing and Neural Networks (ISSIPNN-94), vol. 1, 13–16 April, IEEE Press, Piscataway, NJ, 1994.
[15] Hsien-Wen Tseng, Chin-Chen Chang, " A Very Low Bit Rate Image Compressor Using Transformed Classified Vector Quantization", Informatica ,29 (2005) ,pp. 335–341.
[16] Lenni Yulianti and Tati R.Mengko, "Application of Hybrid Fractal Image Compression Method for Aerial Photographs", *MVA2OOO* IAPR Workshop on Machine Vision Applications, Nov. 28-30.2000,pp. 574-577.
[17] S.Parveen Banu, Dr.Y.Venkataramani,"An Efficient Hybrid Image Compression Scheme based on Correlation of Pixels for Storage and Transmission of Images",International Journal of Computer Applications (0975 – 8887) Volume 18– No.3, 2011,pp. 6-9.
[18] http://nptel.iitm.ac.in/courses /Webcourse contents / IIT % 20Kharagpur /Multimedia % 20Processing /pdf ssg_m6l16.pdf.
[19]GRGIC *et al.*, "Performance analysis of image compression using wavelets", IEEE Trans. Indus.Elect., Vol. 48, No. 3, June 2001,pp. 682-695.
[20]Kamrul Hasan Talukder and Koichi Harada,"Haar Wavelet Based Approach for Image Compression and Quality Assessment of Compressed Image", International Journal of Applied Mathematics, 36:1, IJAM_36_1_9,2007.
[21] Rene J. van der Vleuten, Richard P.Kleihorstt, Christian Hentschel ,"Low-Complexity Scalable DCT Image Compression", 2000 IEEE.
[22] S. Smoot, L. Rowe, "Study of DCT coefficient distributions", Proceedings of the SPIE Symposium on Electronic Imaging, vol. 2657, SPIE, Bellingham, WA, 1996.
[23] D. Monro, B. Sherlock, "Optimum DCT quantization", Proceedings of the IEEE Data Compression Conference (DCC-93), IEEE Press, Piscataway, NJ, 1993.
[24] F. Muller, "Distribution shape of two-dimensional DCT coefficients of natural images", Electron. Lett. 29 (October (22)) (1993), pp. 1935–1936.
[25] Hong, S. W. Bao, P., "Hybrid image compression model based on subband coding and edge preserving regularization", Vision, Image and Signal Processing, IEE Proceedings, Volume: 147, Issue: 1, Feb 2000,pp. 16-22.
[26] Zhe-Ming Lu, Hui Pei ,"Hybrid Image Compression Scheme Based on PVQ and DCTVQ", IEICE - Transactions on Information and Systems archive, Vol E88-D , Issue 10 ,October 2006.
[27] Charles F. Hall., "A Hybrid Image Compression Technique" CH2I 18-8/85/0000-0149 © 1985, IEEE.
[28] Ms.Mansi Kambli1 and Ms.Shalini Bhatia,"Comparison of different Fingerprint Compression Techniques", Signal & Image Processing: An International Journal(SIPIJ) Vol.1, No.1, September 2010, pp. 27-39.
[29] M.Mohamed Sathik , K.Senthamarai Kannan , Y.Jacob Vetha Raj, "Hybrid JPEG Compression Using Edge Based Segmentation", Signal & Image Processing : An International Journal(SIPIJ) Vol.2, No.1, March 2011,pp.165-176.